\documentclass[useAMS,usenatbib,a4paper,referee]{mn2e}
\usepackage{graphicx,times}
\usepackage{amsmath}
\usepackage{longtable}

\newcommand{\hii}{H\textsc{ii}}
\newcommand{\hb}{\mathrm{H}\beta}
\newcommand{\ha}{\mathrm{H}\alpha}
\newcommand{\Oiii}{\mathrm{O\ III}}

\title[The \hii\ Galaxy Hubble Diagram]
  {The \hii\ Galaxy Hubble Diagram Strongly Favors $R_{\rm h}=ct$ over $\Lambda$CDM}
  \author[Wei, Wu \& Melia]
    {Jun-Jie Wei$^{1,2}$\thanks{Email:jjwei@pmo.ac.cn}, Xue-Feng Wu$^{1,3}$\thanks{Email:xfwu@pmo.ac.cn}, and Fulvio Melia$^{1,4}$\thanks{John Woodruff Simpson Fellow. Email: fmelia@email.arizona.edu} \\
  $^1$Purple Mountain Observatory, Chinese Academy of Sciences, Nanjing 210008, China\\
  $^2$Guangxi Key Laboratory for Relativistic Astrophysics, Nanning 530004, China\\
  $^3$Joint Center for Particle, Nuclear Physics and Cosmology, Nanjing University-Purple Mountain Observatory, Nanjing 210008, China\\
  $^4$Department of Physics, The Applied Math Program, and Department of Astronomy, The University of Arizona, AZ 85721, USA}

\begin{document}


%

\maketitle

\begin{abstract}
We continue to build support for the proposal to use \hii~galaxies (HIIGx)
and giant extragalactic \hii~regions (GEHR) as standard candles to construct
the Hubble diagram at redshifts beyond the current reach of Type Ia supernovae.
Using a  sample of 25 high-redshift HIIGx, 107 local HIIGx, and 24 GEHR, we confirm
that the correlation between the emission-line luminosity and ionized-gas velocity dispersion
is a viable luminosity indicator, and use it to test and compare the standard model
$\Lambda$CDM and the $R_{\rm h}=ct$ Universe by optimizing
the parameters in each cosmology using a maximization of the likelihood function.
For the flat $\Lambda$CDM model, the best fit is obtained with $\Omega_{\rm m}=
0.40_{-0.09}^{+0.09}$. However, statistical tools, such as the Akaike (AIC), Kullback (KIC)
and Bayes (BIC) Information Criteria favor $R_{\rm h}=ct$ over the standard
model with a likelihood of $\approx 94.8\%-98.8\%$ versus only $\approx 1.2\%-5.2\%$.
For $w$CDM (the version of $\Lambda$CDM with a dark-energy equation of
state $w_{\rm de}\equiv p_{\rm de}/\rho_{\rm de}$ rather than $w_{\rm de}=w_{\Lambda}=-1$),
a statistically acceptable fit is realized with $\Omega_{\rm m}=0.22_{-0.14}^{+0.16}$ and $w_{\rm de}=
-0.51_{-0.25}^{+0.15}$ which, however, are not fully consistent with their concordance
values. In this case, $w$CDM has two more free parameters than $R_{\rm h}=ct$,
and is penalized more heavily by these criteria. We find that $R_{\rm h}=ct$ is strongly
favored over $w$CDM with a likelihood of $\approx 92.9\%-99.6\%$ versus
only $0.4\%-7.1\%$. The current HIIGx sample is already large enough for the
BIC to rule out $\Lambda$CDM/$w$CDM in favor of $R_{\rm h}=ct$ at a confidence level
approaching $3\sigma$.
\end{abstract}

\section{Introduction}
\hii~galaxies (HIIGx) are massive and compact aggregates of star formation.
The total luminosity of an HIIGx is almost completely dominated by the starburst.
Giant extragalactic \hii~regions (GEHR) also have massive bursts of star formation, but
are generally located in the outer discs of late-type galaxies. In brief, \hii~galaxies and
the \hii~regions of galaxies are characterized by rapidly forming stars surrounded by
ionized hydrogen, the presence of which leads to their naming convention.
It is well known that HIIGx and GEHR are physically similar systems (Melnick
et al. 1987); indeed, their optical spectra are indistinguishable, and are characterized by
strong Balmer emission lines in $\ha$ and $\hb$ produced by the hydrogen ionized by the
young massive star clusters (Searle \& Sargent 1972; Bergeron 1977; Terlevich
\& Melnick 1981; Kunth \& \"{O}stlin 2000).

Since the starburst component can reach very high luminosities, HIIGx can be detected
at relatively high redshifts ($z > 3$). What really makes these galaxies interesting
as standard candles (e.g., Melnick et al. 2000; Siegel et al. 2005) is the fact
that as the mass of the starburst component increases, both the number of ionizing photons
and the turbulent velocity of the gas, which is dominated by gravitational potential of
the star and gas, also increase. This naturally induces a correlation between the luminosity
$L(\hb)$ in $\hb$ and the ionized gas velocity dispersion $\sigma$ (Terlevich
\& Melnick 1981). The scatter in this relation is small enough that it can be used as
a cosmic distance indicator independently of redshift (see Melnick et al. 1987; Melnick
et al. 1988; Fuentes-Masip et al. 2000; Melnick et al. 2000; Bosch et al. 2002;
Telles 2003; Siegel et al. 2005; Bordalo \& Telles 2011; Plionis et al. 2011;
Mania \& Ratra 2012; Ch{\'a}vez et al. 2012, 2014; Terlevich et al. 2015).

With HIIGx and GEHR as local calibrators, the first attempt to determine the Hubble constant
$H_0$ was presented in Melnick et al. (1988). Ch{\'a}vez et al. (2012) subsequently provided
accurate estimates of $H_0$ using the $L(\hb)-\sigma$ correlation for GEHR and local HIIGx.
The use of intermediate and high-$z$ HIIGx as deep cosmological tracers was
discussed by Melnick et al. (2000), who confirmed that the $L(\hb)-\sigma$ correlation
is valid for high-redshift HIIGx up to $z\sim3$. Siegel et al. (2005) used a sample
of 15 high-$z$ HIIGx ($2.17<z<3.39$) to constrain the normalized mass density $\Omega_{\rm m}$,
producing a best-fitting value of $\Omega_{\rm m}=0.21^{+0.30}_{-0.12}$ for a flat $\Lambda$CDM
cosmology. This analysis was extended by Plionis et al. (2011), who investigated the viability of
using HIIGx to constrain the dark energy equation of state, and showed that the HIIGx
$L(\hb)-\sigma$ correlation is a viable high-$z$ tracer. Using the biggest sample
to date (156 combined sources, including 25 high-$z$ HIIGx, 107 local HIIGx, and 24 GEHR),
Terlevich et al. (2015) were able to constrain the cosmological parameters, showing
that they are consistent with the analysis of Type Ia supernovae.

In this paper, we will use the newer and larger sample of HIIGx from Terlevich et al. (2015)
to examine whether the HIIGx can be utilized---not only to optimize the parameters in $\Lambda$CDM
(e.g., Siegel et al. 2005; Plionis et al. 2011; Terlevich et al. 2015)---but
also to carry out comparative studies between competing cosmologies,
such as $\Lambda$CDM and the $R_{\rm h}=ct$ Universe (Melia 2003, 2007, 2013a, 2016a, 2016b;
Melia \& Abdelqader 2009; Melia \& Shevchuk 2012). Like $\Lambda$CDM, the $R_{\rm h}=ct$
Universe is a Friedmann-Robertson-Walker (FRW) cosmology that assumes the presence
of dark energy, as well as matter and radiation. The principle difference between them is
that the latter is also constrained by the equation of state $\rho+3p=0$ (the so-called
zero active mass condition in general relativity; Melia 2016a, 2016b), in terms of the
total pressure $p$ and energy density $\rho$.

An examination of which of these two models, $\Lambda$CDM or $R_{\rm h}=ct$,
is favoured by the observations has been carried out using a diverse range of data
over a period of more than 10 years. These observations include high-$z$ quasars
(e.g., Kauffmann \& Haehnelt 2000; Wyithe \& Loeb 2003; Melia 2013b, 2014; Melia
\& McClintock 2015b), Gamma-ray bursts (e.g., Dai et al. 2004; Ghirlanda et al.
2004; Wei et al. 2013), cosmic chronometers (e.g., Jimenez \& Loeb 2002; Simon
et al. 2005; Melia \& Maier 2013; Melia \& McClintock 2015a), Type Ia supernovae
(e.g., Perlmutter et al. 1998; Riess et al. 1998; Schmidt et al. 1998; Melia 2012;
Wei et al. 2015b), Type Ic superluminous supernovae (e.g., Inserra \& Smart 2014;
Wei et al. 2015a), and the age measurements of passively evolving galaxies (e.g.,
Alcaniz \& Lima 1999; Lima \& Alcaniz 2000; Wei et al. 2015c). In all such one-on-one
comparisons completed thus far, model selection tools show that the data favour
$R_{\rm h}=ct$ over $\Lambda$CDM (see, e.g., Melia 2013b, 2014; Melia \& Maier 2013;
Melia \& McClintock 2015a, 2015b; Wei et al. 2013, 2015a, 2015b, 2015c).

In this paper, we extend the comparison between $R_{\rm h}=ct$ and
$\Lambda$CDM by now including HIIGx in this study. In \S~2, we
will briefly describe the currently available sample and our method
of analysis, and then constrain the cosmological parameters---both
in the context of $\Lambda$CDM and the $R_{\rm h}=ct$ universe (\S~3).
In \S~4, we will construct the \hii\ Galaxy Hubble diagrams for these
two expansion scenarios, and discuss the model selection tools we use to
test them. We end with our conclusions in \S~5.

\vfill\newpage
\section{Observational data and methodology}
A total sample of 156 sources (25 high-$z$ \hii~galaxies, 107 local \hii~galaxies,
and 24 giant extragalactic \hii~regions) assembled by Terlevich et al. (2015) are
appropriate for this work, and we base our analysis on the methodology described in
their paper.

A catalog of 128 local \hii~galaxies was selected from the SDSS DR7 spectroscopic
catalogue (Abazajian et al. 2009) for having the strongest Balmer emission lines
relative to the continuum (i.e., the largest equivalent width, $EW(\hb)>50$\AA,
in their $\hb$ emission
lines) and in the redshift range $\sim0.01<z<0.2$ (Ch{\'a}vez et al. 2014).
The lower limit of the equivalent width of $\hb$ was selected to avoid starbursts that
are either evolved or contaminated by an underlying older stellar population component
(e.g., Melnick et al. 2000). The lower redshift limit was set to avoid nearby objects
that are more affected by local peculiar motions relative to the Hubble flow and
the upper limit was chosen to minimize any possible Malmquist bias and to avoid gross
cosmological effects. From this observed sample, Ch{\'a}vez et al. (2014) removed
13 objects with a low S/N or that showed evidence for a prominent underlying Balmer
absorption. They also removed an extra object with highly asymmetric emission
lines. After this cut, 114 objects were left that comprise their
`initial' sample. Melnick et al. (1988) showed that imposing an upper limit
to the velocity dispersion, such as $\log \sigma(\hb)<1.8$ km $\rm s^{-1}$, minimizes
the probability of including rotationally supported systems
and/or objects with multiple young ionizing clusters contributing to
the total flux and affecting the line profiles. Therefore, they
selected all objects having $\log \sigma(\hb)<1.8$ km $\rm s^{-1}$
from the `initial' sample, thus creating their `benchmark' catalog comprised of 107 local objects.

Following the same sample selection criteria, Terlevich et al. (2015)
presented observations of a sample of 6 high-$z$ HIIGx in the redshift
range of $0.64\leq z \leq2.33$ obtained with the XShooter spectrograph
at the Cassegrain focus of the ESO-VLT (European Southern Observatory Very Large Telescope).
The addition of 19 high-$z$ HIIGx from the literature---6 HIIGx from Erb
et al. (2006a,b), 1 from Maseda et al. (2014) and 12 from Masters et al.
(2014)---yields the total set of 25 high-$z$ HIIGx. Ch{\'a}vez et al.
(2012) first gathered the necessary data from the literature to compile a sample
of 24 GEHR in nine nearby galaxies. For these objects, the velocity dispersions
and the global integrated $\hb$ fluxes with corresponding extinction were
taken from Melnick et al. (1987). In summary, our sample contains 156 objects,
whose properties are summarized in Table~1.

\begin{center}
\begin{small}
\begin{longtable}{lcccc}
\caption[GRB Catalog.]{Flux and gas velocity dispersion of \hii~Galaxies and Giant \hii\ regions.} \label{table} \\
\hline
Name&\emph{z}&$\log \sigma(\hb)$&$\log F(\hb)$&Ref.\\
&  & High-$z$ \hii~Galaxies  &  & \\
\hline
Q2343-BM133	&	1.47740	&	1.756	$\pm$	0.017	&	-15.884	$\pm$	0.043	&	1	\\
Q2343-BX418	&	2.30520	&	1.758	$\pm$	0.016	&	-16.518	$\pm$	0.017	&	1	\\
Q2343-BX660	&	2.17350	&	1.808	$\pm$	0.016	&	-16.473	$\pm$	0.019	&	2	\\
HoyosD2-5	&	0.63640	&	1.597	$\pm$	0.023	&	-15.791	$\pm$	0.177	&	2	\\
HoyosD2-1	&	0.85100	&	1.695	$\pm$	0.049	&	-15.801	$\pm$	0.177	&	2	\\
HoyosD2-12	&	0.68160	&	1.527	$\pm$	0.027	&	-15.960	$\pm$	0.175	&	1	\\
HDF-BX1277	&	2.27130	&	1.799	$\pm$	0.062	&	-16.637	$\pm$	0.038	&	1	\\
Q0201-B13	&	2.16630	&	1.792	$\pm$	0.070	&	-17.073	$\pm$	0.018	&	1	\\
Q1623-BX215	&	2.18140	&	1.845	$\pm$	0.093	&	-16.641	$\pm$	0.055	&	1	\\
Q1623-BX453	&	2.18160	&	1.785	$\pm$	0.028	&	-16.042	$\pm$	0.099	&	1	\\
Q2346-BX120	&	2.26640	&	1.792	$\pm$	0.084	&	-16.727	$\pm$	0.025	&	1	\\
Q2346-BX405	&	2.03000	&	1.699	$\pm$	0.035	&	-16.300	$\pm$	0.007	&	3	\\
COSMOS-17839	&	1.41200	&	1.664	$\pm$	0.084	&	-16.832	$\pm$	0.427	&	4	\\
WISP159-134	&	1.30000	&	1.686	$\pm$	0.045	&	-16.264	$\pm$	0.042	&	4	\\
WISP173-205	&	1.44400	&	1.834	$\pm$	0.045	&	-16.377	$\pm$	0.055	&	4	\\
WISP46-75	&	1.50400	&	1.839	$\pm$	0.066	&	-16.273	$\pm$	0.146	&	4	\\
WISP22-216	&	1.54300	&	1.641	$\pm$	0.040	&	-16.475	$\pm$	0.045	&	4	\\
WISP64-2056	&	1.61000	&	1.746	$\pm$	0.039	&	-16.461	$\pm$	0.038	&	4	\\
WISP138-173	&	2.15800	&	1.814	$\pm$	0.040	&	-16.372	$\pm$	0.052	&	4	\\
WISP64-210	&	2.17700	&	1.830	$\pm$	0.039	&	-16.456	$\pm$	0.041	&	4	\\
WISP204-133	&	2.19100	&	1.765	$\pm$	0.063	&	-16.899	$\pm$	0.043	&	4	\\
WISP70-253	&	2.21500	&	1.628	$\pm$	0.041	&	-16.927	$\pm$	0.017	&	4	\\
WISP96-158	&	2.23400	&	1.702	$\pm$	0.043	&	-16.562	$\pm$	0.041	&	4	\\
WISP138-160	&	2.26400	&	1.838	$\pm$	0.044	&	-16.223	$\pm$	0.037	&	4	\\
WISP206-261	&	2.31500	&	1.693	$\pm$	0.044	&	-16.411	$\pm$	0.171	&	4	\\
\hline
&  & Local \hii~Galaxies  &  & \\
\hline
J001647-104742	&	0.02203	&	1.377	$\pm$	0.039	&	-13.096	$\pm$	0.141	&	5	\\
J002339-094848	&	0.05191	&	1.463	$\pm$	0.036	&	-13.411	$\pm$	0.120	&	5	\\
J002425+140410	&	0.01257	&	1.538	$\pm$	0.034	&	-13.229	$\pm$	0.049	&	5	\\
J003218+150014	&	0.01636	&	1.577	$\pm$	0.034	&	-13.308	$\pm$	0.086	&	5	\\
J005147+000940	&	0.03637	&	1.454	$\pm$	0.036	&	-13.932	$\pm$	0.049	&	5	\\
J005602-101009	&	0.05712	&	1.529	$\pm$	0.034	&	-13.954	$\pm$	0.098	&	5	\\
J013258-085337	&	0.09424	&	1.527	$\pm$	0.033	&	-14.119	$\pm$	0.049	&	5	\\
J013344+005711	&	0.01812	&	1.283	$\pm$	0.042	&	-13.987	$\pm$	0.062	&	5	\\
J014137-091435	&	0.01718	&	1.369	$\pm$	0.040	&	-13.610	$\pm$	0.109	&	5	\\
J014707+135629	&	0.05574	&	1.625	$\pm$	0.033	&	-13.603	$\pm$	0.075	&	5	\\
J021852-091218	&	0.01207	&	1.144	$\pm$	0.060	&	-13.724	$\pm$	0.120	&	5	\\
J022037-092907	&	0.11235	&	1.706	$\pm$	0.033	&	-13.671	$\pm$	0.075	&	5	\\
J024052-082827	&	0.08164	&	1.651	$\pm$	0.034	&	-13.554	$\pm$	0.075	&	5	\\
J024453-082137	&	0.07687	&	1.590	$\pm$	0.034	&	-13.822	$\pm$	0.062	&	5	\\
J025426-004122	&	0.01420	&	1.390	$\pm$	0.038	&	-13.575	$\pm$	0.049	&	5	\\
J030321-075923	&	0.16417	&	1.782	$\pm$	0.032	&	-14.041	$\pm$	0.049	&	5	\\
J031023-083432	&	0.05097	&	1.419	$\pm$	0.039	&	-14.025	$\pm$	0.062	&	5	\\
J033526-003811	&	0.02282	&	1.350	$\pm$	0.041	&	-13.757	$\pm$	0.086	&	5	\\
J040937-051805	&	0.07443	&	1.548	$\pm$	0.034	&	-13.934	$\pm$	0.062	&	5	\\
J051519-391741	&	0.05041	&	1.446	$\pm$	0.026	&	-13.505	$\pm$	0.255	&	5	\\
J074806+193146	&	0.06347	&	1.576	$\pm$	0.025	&	-13.635	$\pm$	0.109	&	5	\\
J074947+154013	&	0.07485	&	1.567	$\pm$	0.022	&	-14.009	$\pm$	0.098	&	5	\\
J080000+274642	&	0.03993	&	1.484	$\pm$	0.026	&	-13.653	$\pm$	0.075	&	5	\\
J080619+194927	&	0.07051	&	1.791	$\pm$	0.032	&	-13.187	$\pm$	0.062	&	5	\\
J081334+313252	&	0.02021	&	1.463	$\pm$	0.035	&	-13.331	$\pm$	0.098	&	5	\\
J081403+235328	&	0.02077	&	1.480	$\pm$	0.026	&	-13.755	$\pm$	0.075	&	5	\\
J081420+575008	&	0.05547	&	1.565	$\pm$	0.033	&	-13.808	$\pm$	0.062	&	5	\\
J081737+520236	&	0.02370	&	1.588	$\pm$	0.033	&	-13.230	$\pm$	0.141	&	5	\\
J082520+082723	&	0.08769	&	1.532	$\pm$	0.035	&	-14.116	$\pm$	0.109	&	5	\\
J082530+504804	&	0.09729	&	1.649	$\pm$	0.033	&	-13.736	$\pm$	0.062	&	5	\\
J082722+202612	&	0.10937	&	1.688	$\pm$	0.025	&	-13.538	$\pm$	0.120	&	5	\\
J083946+140033	&	0.11245	&	1.707	$\pm$	0.024	&	-13.852	$\pm$	0.086	&	5	\\
J084000+180531	&	0.07302	&	1.664	$\pm$	0.019	&	-13.767	$\pm$	0.062	&	5	\\
J084029+470710	&	0.04258	&	1.637	$\pm$	0.034	&	-13.169	$\pm$	0.086	&	5	\\
J084219+300703	&	0.08479	&	1.652	$\pm$	0.024	&	-13.587	$\pm$	0.086	&	5	\\
J084220+115000	&	0.03065	&	1.490	$\pm$	0.035	&	-13.113	$\pm$	0.120	&	5	\\
J084414+022621	&	0.09209	&	1.747	$\pm$	0.024	&	-13.289	$\pm$	0.109	&	5	\\
J084527+530852	&	0.03127	&	1.449	$\pm$	0.035	&	-13.451	$\pm$	0.086	&	5	\\
J084634+362620	&	0.01125	&	1.406	$\pm$	0.040	&	-13.023	$\pm$	0.098	&	5	\\
J085221+121651	&	0.07687	&	1.725	$\pm$	0.032	&	-13.152	$\pm$	0.075	&	5	\\
J090418+260106	&	0.09922	&	1.766	$\pm$	0.024	&	-13.513	$\pm$	0.086	&	5	\\
J090506+223833	&	0.12641	&	1.646	$\pm$	0.025	&	-13.894	$\pm$	0.062	&	5	\\
J090531+033530	&	0.04038	&	1.566	$\pm$	0.025	&	-13.763	$\pm$	0.049	&	5	\\
J091434+470207	&	0.02771	&	1.535	$\pm$	0.035	&	-13.156	$\pm$	0.033	&	5	\\
J091640+182807	&	0.02293	&	1.477	$\pm$	0.035	&	-13.651	$\pm$	0.075	&	5	\\
J091652+003113	&	0.05815	&	1.614	$\pm$	0.024	&	-13.919	$\pm$	0.086	&	5	\\
J092749+084037	&	0.10809	&	1.737	$\pm$	0.023	&	-13.668	$\pm$	0.120	&	5	\\
J092918+002813	&	0.09494	&	1.561	$\pm$	0.025	&	-13.915	$\pm$	0.086	&	5	\\
J093006+602653	&	0.01352	&	1.441	$\pm$	0.036	&	-13.232	$\pm$	0.049	&	5	\\
J093424+222522	&	0.08536	&	1.700	$\pm$	0.024	&	-13.693	$\pm$	0.075	&	5	\\
J093813+542825	&	0.10263	&	1.787	$\pm$	0.031	&	-13.343	$\pm$	0.062	&	5	\\
J094000+203122	&	0.04587	&	1.602	$\pm$	0.025	&	-14.003	$\pm$	0.033	&	5	\\
J094252+354725	&	0.01558	&	1.536	$\pm$	0.034	&	-13.485	$\pm$	0.049	&	5	\\
J094254+340411	&	0.02329	&	1.496	$\pm$	0.036	&	-14.095	$\pm$	0.086	&	5	\\
J094809+425713	&	0.01765	&	1.434	$\pm$	0.036	&	-13.534	$\pm$	0.049	&	5	\\
J095000+300341	&	0.01822	&	1.440	$\pm$	0.035	&	-13.541	$\pm$	0.075	&	5	\\
J095023+004229	&	0.09883	&	1.750	$\pm$	0.025	&	-13.640	$\pm$	0.086	&	5	\\
J095226+021759	&	0.12029	&	1.746	$\pm$	0.025	&	-13.591	$\pm$	0.086	&	5	\\
J095227+322809	&	0.01578	&	1.296	$\pm$	0.044	&	-13.516	$\pm$	0.062	&	5	\\
J095545+413429	&	0.01621	&	1.425	$\pm$	0.042	&	-13.220	$\pm$	0.109	&	5	\\
J100720+193349	&	0.03259	&	1.297	$\pm$	0.035	&	-14.176	$\pm$	0.086	&	5	\\
J100746+025228	&	0.02518	&	1.532	$\pm$	0.034	&	-13.372	$\pm$	0.086	&	5	\\
J101042+125516	&	0.06244	&	1.681	$\pm$	0.042	&	-13.261	$\pm$	0.049	&	5	\\
J101136+263027	&	0.05564	&	1.612	$\pm$	0.025	&	-13.661	$\pm$	0.086	&	5	\\
J101157+130822	&	0.14486	&	1.709	$\pm$	0.032	&	-13.922	$\pm$	0.062	&	5	\\
J101430+004755	&	0.14807	&	1.774	$\pm$	0.024	&	-13.811	$\pm$	0.062	&	5	\\
J101458+193219	&	0.01390	&	1.279	$\pm$	0.044	&	-13.986	$\pm$	0.075	&	5	\\
J102429+052451	&	0.03476	&	1.560	$\pm$	0.037	&	-13.172	$\pm$	0.049	&	5	\\
J102732-284201	&	0.03375	&	1.540	$\pm$	0.034	&	-13.347	$\pm$	0.296	&	5	\\
J103328+070801	&	0.04583	&	1.791	$\pm$	0.033	&	-12.873	$\pm$	0.109	&	5	\\
J103412+014249	&	0.06988	&	1.597	$\pm$	0.022	&	-13.999	$\pm$	0.098	&	5	\\
J103509+094516	&	0.05050	&	1.630	$\pm$	0.034	&	-14.027	$\pm$	0.062	&	5	\\
J103726+270759	&	0.07806	&	1.593	$\pm$	0.025	&	-13.865	$\pm$	0.086	&	5	\\
J104457+035313	&	0.01453	&	1.410	$\pm$	0.038	&	-13.365	$\pm$	0.049	&	5	\\
J104554+010405	&	0.02777	&	1.593	$\pm$	0.034	&	-13.037	$\pm$	0.075	&	5	\\
J104653+134645	&	0.01216	&	1.446	$\pm$	0.036	&	-13.260	$\pm$	0.062	&	5	\\
J104723+302144	&	0.03039	&	1.639	$\pm$	0.033	&	-12.656	$\pm$	0.109	&	5	\\
J105032+153806	&	0.08564	&	1.561	$\pm$	0.033	&	-13.326	$\pm$	0.062	&	5	\\
J105040+342947	&	0.05314	&	1.544	$\pm$	0.026	&	-13.561	$\pm$	0.062	&	5	\\
J105108+131927	&	0.04670	&	1.569	$\pm$	0.034	&	-13.909	$\pm$	0.075	&	5	\\
J105210+032713	&	0.15134	&	1.587	$\pm$	0.032	&	-14.180	$\pm$	0.075	&	5	\\
J105331+011740	&	0.12499	&	1.660	$\pm$	0.024	&	-13.974	$\pm$	0.062	&	5	\\
J105741+653539	&	0.01111	&	1.396	$\pm$	0.038	&	-13.372	$\pm$	0.086	&	5	\\
J110838+223809	&	0.02492	&	1.434	$\pm$	0.027	&	-13.463	$\pm$	0.062	&	5	\\
J121329+114056	&	0.02187	&	1.465	$\pm$	0.016	&	-13.360	$\pm$	0.086	&	5	\\
J121717-280233	&	0.02765	&	1.407	$\pm$	0.020	&	-13.364	$\pm$	0.275	&	5	\\
J131235+125743	&	0.02671	&	1.431	$\pm$	0.022	&	-13.434	$\pm$	0.120	&	5	\\
J132347-013252	&	0.02362	&	1.309	$\pm$	0.032	&	-13.567	$\pm$	0.062	&	5	\\
J132549+330354	&	0.01508	&	1.424	$\pm$	0.027	&	-13.407	$\pm$	0.049	&	5	\\
J134531+044232	&	0.03138	&	1.609	$\pm$	0.024	&	-13.234	$\pm$	0.062	&	5	\\
J142342+225728	&	0.03328	&	1.683	$\pm$	0.024	&	-13.475	$\pm$	0.120	&	5	\\
J144805-011057	&	0.02808	&	1.688	$\pm$	0.024	&	-12.907	$\pm$	0.062	&	5	\\
J162152+151855	&	0.03437	&	1.739	$\pm$	0.023	&	-13.263	$\pm$	0.005	&	5	\\
J171236+321633	&	0.01094	&	1.340	$\pm$	0.021	&	-13.608	$\pm$	0.062	&	5	\\
J192758-413432	&	0.00880	&	1.494	$\pm$	0.025	&	-12.579	$\pm$	0.235	&	5	\\
J211527-075951	&	0.02711	&	1.397	$\pm$	0.017	&	-13.537	$\pm$	0.075	&	5	\\
J212332-074831	&	0.02662	&	1.441	$\pm$	0.036	&	-13.861	$\pm$	0.098	&	5	\\
J214350-072003	&	0.10880	&	1.559	$\pm$	0.046	&	-14.114	$\pm$	0.075	&	5	\\
J220802+131334	&	0.11506	&	1.757	$\pm$	0.033	&	-13.772	$\pm$	0.120	&	5	\\
J221823+003918	&	0.10726	&	1.707	$\pm$	0.025	&	-13.911	$\pm$	0.141	&	5	\\
J222510-001152	&	0.06551	&	1.627	$\pm$	0.031	&	-13.653	$\pm$	0.062	&	5	\\
J224556+125022	&	0.07928	&	1.662	$\pm$	0.033	&	-13.499	$\pm$	0.086	&	5	\\
J225140+132713	&	0.06094	&	1.660	$\pm$	0.033	&	-13.120	$\pm$	0.086	&	5	\\
J230117+135230	&	0.02283	&	1.318	$\pm$	0.046	&	-13.527	$\pm$	0.098	&	5	\\
J230123+133314	&	0.02873	&	1.568	$\pm$	0.033	&	-13.147	$\pm$	0.098	&	5	\\
J231442+010621	&	0.03278	&	1.393	$\pm$	0.041	&	-14.091	$\pm$	0.086	&	5	\\
J232936-011056	&	0.06479	&	1.573	$\pm$	0.033	&	-13.723	$\pm$	0.098	&	5	\\
\hline
&  & Giant \hii\ regions  &  & \\
\hline
GEHR	&	0.00012	&	1.013	$\pm$	0.035	&	-11.131	$\pm$	0.102	&	6	\\
GEHR	&	0.00012	&	1.021	$\pm$	0.035	&	-11.137	$\pm$	0.095	&	6	\\
GEHR	&	0.00001	&	1.061	$\pm$	0.035	&	-9.083	$\pm$	0.095	&	6	\\
GEHR	&	0.00020	&	1.111	$\pm$	0.035	&	-11.269	$\pm$	0.095	&	6	\\
GEHR	&	0.00110	&	1.133	$\pm$	0.036	&	-12.509	$\pm$	0.102	&	6	\\
GEHR	&	0.00110	&	1.159	$\pm$	0.035	&	-12.181	$\pm$	0.102	&	6	\\
GEHR	&	0.00085	&	1.176	$\pm$	0.035	&	-11.953	$\pm$	0.102	&	6	\\
GEHR	&	0.00100	&	1.199	$\pm$	0.035	&	-12.185	$\pm$	0.095	&	6	\\
GEHR	&	0.00077	&	1.204	$\pm$	0.035	&	-12.101	$\pm$	0.095	&	6	\\
GEHR	&	0.00020	&	1.201	$\pm$	0.035	&	-11.082	$\pm$	0.095	&	6	\\
GEHR	&	0.00020	&	1.250	$\pm$	0.036	&	-10.733	$\pm$	0.102	&	6	\\
GEHR	&	0.00100	&	1.250	$\pm$	0.036	&	-12.232	$\pm$	0.095	&	6	\\
GEHR	&	0.00185	&	1.267	$\pm$	0.035	&	-12.619	$\pm$	0.095	&	6	\\
GEHR	&	0.00085	&	1.207	$\pm$	0.035	&	-11.571	$\pm$	0.095	&	6	\\
GEHR	&	0.00077	&	1.267	$\pm$	0.035	&	-11.579	$\pm$	0.095	&	6	\\
GEHR	&	0.00020	&	1.277	$\pm$	0.035	&	-10.285	$\pm$	0.095	&	6	\\
GEHR	&	0.00185	&	1.293	$\pm$	0.035	&	-12.278	$\pm$	0.102	&	6	\\
GEHR	&	0.00077	&	1.320	$\pm$	0.035	&	-11.713	$\pm$	0.102	&	6	\\
GEHR	&	0.00001	&	1.369	$\pm$	0.035	&	-7.959	$\pm$	0.095	&	6	\\
GEHR	&	0.00077	&	1.384	$\pm$	0.035	&	-11.258	$\pm$	0.102	&	6	\\
GEHR	&	0.00185	&	1.314	$\pm$	0.035	&	-11.983	$\pm$	0.095	&	6	\\
GEHR	&	0.00185	&	1.310	$\pm$	0.035	&	-11.775	$\pm$	0.095	&	6	\\
GEHR	&	0.00185	&	1.333	$\pm$	0.035	&	-11.695	$\pm$	0.095	&	6	\\
GEHR	&	0.00185	&	1.351	$\pm$	0.035	&	-11.722	$\pm$	0.095	&	6	\\		
\hline
\end{longtable}
\footnotesize
Reference: (1) Erb et al. 2006b; (2) Hoyos et al. 2005;
(3) Maseda et al. 2014;\\ (4) Masters et al. 2014; (5) Ch{\'a}vez et al. 2014; (6) Terlevich et al. 2015.
\end{small}
\end{center}

The observed velocity dispersions ($\sigma_{0}$) and their $1\sigma$ uncertainties
were derived from the full width at half-maximum (FWHM) measurements of the
$\hb$ and $[\Oiii]\lambda5007$ lines, i.e., $\sigma_{0}\equiv\frac{FWHM}{2\sqrt{2\ln(2)}}$.
The values of $\sigma_{0}$ were corrected for thermal ($\sigma_{\rm th}$), instrumental
($\sigma_{\rm i}$) and fine-structure ($\sigma_{\rm fs}$) broadening, yielding a final
corrected velocity dispersion
\begin{equation}
\sigma=\sqrt{\sigma_{0}^{2}-\sigma_{\rm th}^{2}-\sigma_{\rm i}^{2}-\sigma_{\rm fs}^{2}}\;.
\end{equation}
We adopted a value of $\sigma_{\rm fs}(\hb)=2.4$ km $\rm s^{-1}$ from Garc{\'{\i}}a-D{\'{\i}}az et al. (2008).
A detailed discussion of the other terms in this equation can be found in Ch{\'a}vez et al. (2014).
The corrected emission line velocity dispersions and their $1\sigma$ uncertainties
are shown in Table~1, column~(3).

$\hb$ integrated fluxes can be measured by fitting a single Gaussian to the long-slit spectra.
Terlevich et al. (2015) adopted the reddening corrections from the literature, where
the extinction, $A_{v}$, was derived from the published $E(B-V)$ using a standard reddening curve
with $R_{v}=A_{v}/E(B-V)=4.05$ (Calzetti et al. 2000). For those objects where the reddening
corrections were not available, the mean $A_{v}=0.33$ from the local HIIGx was adopted.
The reddening corrected $\hb$ fluxes and their $1\sigma$ uncertainties are shown in Table~1, column~(4).
With the data listed in Table~1, the $\hb$ luminosity can be calculated from the expression
\begin{equation}
 L(\hb) = 4 \pi  D_L^2(z) F(\hb)\;,
\end{equation}
where $D_L$ is the cosmology-dependent luminosity distance at redshift $z$
and $F(\mathrm{H}\beta)$ is the reddening corrected $\hb$ flux.

The emission-line luminosity versus ionized gas velocity dispersion ($L - \sigma$) correlation
is (Ch{\'a}vez et al. 2012; Ch{\'a}vez et al. 2014; Terlevich et al. 2015)
\begin{equation}
\log L(\hb)=\alpha \log \sigma(\hb)+\kappa\;,
\end{equation}
where $\alpha$ is the slope and $\kappa$ is a constant representing
the logarithmic luminosity at $\log \sigma(\hb)=0$.
The scatter of the empirical correlation for $L(\hb)$ is so small that it
has been used as a luminosity indicator for cosmology (e.g., Ch{\'a}vez et al. 2012;
Terlevich et al. 2015). However, since this luminosity indicator is cosmology-dependent,
we cannot use it to constrain the cosmological parameters directly. In order to
avoid circularity issues, the coefficients $\alpha$ and $\kappa$ must be optimized
simultaneously with the cosmological parameters.

With the $L - \sigma$ relation, the distance modulus of an \hii~galaxy can be
obtained as
\begin{equation}
\mu_{\rm obs}=2.5\left[\kappa +\alpha \log \sigma(\hb) - \log F(\hb)\right]-100.2 \;.
\end{equation}
The error $\sigma_{\mu_{\rm obs}}$ in $\mu_{\rm obs}$ is calculated by error propagation,
i.e.,
\begin{equation}\label{sigma}
\sigma_{\mu_{\rm obs}}=2.5\left[\left(\alpha \sigma_{\log \sigma}\right)^{2}+\left(\sigma_{\log F}\right)^{2}\right]^{1/2} \;,
\end{equation}
where $\sigma_{\log \sigma}$ and $\sigma_{\log F}$ are the $1\sigma$ uncertainties in $\log \sigma(\hb)$
and $\log F(\hb)$, respectively.

The theoretical distance modulus $\mu_{\rm th}$ of an \hii~galaxy at redshift $z$ is defined as
\begin{equation}
\mu_{\rm th}\equiv5 \log\left[\frac{D_{L}(z)}{\rm Mpc}\right]+25\;,
\end{equation}
in terms of the luminosity distance $D_L$, whose determination requires the assumption
of a particular expansion scenario.  Both $\Lambda$CDM and $R_{\rm h}=ct$ are FRW
cosmologies, but to calculate the expansion rate, one needs to assume
for the former specific constituents in the density,
written as $\rho=\allowbreak\rho_r+\nobreak\rho_m+\nobreak\rho_{\rm de}$,
where $\rho_r$, $\rho_m$, and $\rho_{\rm de}$ are, respectively, the energy densities
of radiation, matter (luminous and dark), and dark energy.
These are often expressed in~terms of today's critical
density, $\rho_c\equiv 3c^2 H_0^2/8\pi G$, where $H_0$~is the Hubble
constant, by $\Omega_m\equiv\rho_m/\rho_c$, $\Omega_r\equiv\rho_r/\rho_c$,
and $\Omega_{\rm de} \equiv \rho_{\rm de}/\rho_c$.  In~a flat universe with
zero spatial curvature, the total scaled energy density is
$\Omega\equiv\allowbreak\Omega_m+\nobreak\Omega_r+\nobreak\Omega_{\rm de}\nobreak=\nobreak1$.
In $R_{\rm h}=ct$, on the other hand, whatever constituents are present
in~$\rho$, the principal constraint is the total equation-of-state
$p=-\rho/3$.

In $\Lambda$CDM, the luminosity distance is given as
\begin{equation}\label{DL_LCDM}
D_{L}^{\Lambda {\rm CDM}}(z) = {c\over
  H_{0}}\,{(1+z)\over\sqrt{\mid\Omega_{k}\mid}}\; {\rm
  sinn}\Biggl\{\mid\Omega_{k}\mid^{1/2}
\int_{0}^{z}{dz\over\sqrt{\Omega_{\rm m}(1+z)^{3}+\Omega_{k}(1+z)^{2}+\Omega_{\rm de}(1+z)^{3(1+w_{\rm de})}}}\Biggr\}\;,
\end{equation}
where $p_{\rm de}=w_{\rm de}\rho_{\rm de}$ is the dark-energy equation
of state, and we have assumed that the radiation density is negligible in the local Universe. Also,
$\Omega_{k}=1-\Omega_{\rm m}-\Omega_{\rm de}$ represents the spatial curvature of the
Universe---appearing as a term proportional to the spatial curvature
constant $k$ in the Friedmann equation.  In addition, sinn is $\sinh$ when
$\Omega_{k}>0$ and $\sin$ when $\Omega_{k}<0$.  For a flat Universe
($\Omega_{k}=0$), the right side becomes $(1+z)c/H_{0}$ times the
indefinite integral.

In the $R_{\rm h}=ct$ Universe (Melia 2003, 2007, 2013a, 2016a, 2016b; Melia \& Abdelqader 2009;
Melia \& Shevchuk 2012), the luminosity distance is given by the much simpler expression
\begin{equation}\label{DL_Rh}
D_L^{R_{\rm h}=ct}(z)={c\over H_0}(1+z)\ln(1+z)\;.
\end{equation}

To find the best-fit cosmological parameters and (simultaneously) the coefficients $\alpha$ and $\kappa$,
we adopt the method of maximum likelihood estimation (MLE; see Wei et al. 2015b). The joint likelihood function
for all these parameters, based on a flat Bayesian prior, is
\begin{equation}\label{likelihood1}
\mathcal{L} = \prod_{i}
\frac{1}{\sqrt{2\pi}\,\sigma_{\mu_{\rm obs}, i}}\;\times
\exp\left[-\,\frac{\left(\mu_{{\rm obs},i}-\mu_{\rm
      th}(z_i)\right)^{2}}{2\sigma^{2}_{\mu_{\rm obs}, i}}\right] .
\end{equation}
Because the first factor $\sigma_{\mu_{\rm obs}}$ in the product of Equation~(\ref{likelihood1}) is not
a constant, dependent on the value of $\alpha$ (see Equation~\ref{sigma}), maximizing the
likelihood function $\mathcal{L}$ is not exactly equivalent to minimizing the $\chi^{2}$
statistic, i.e., $\chi^{2}=\sum_i\frac{\left(\mu_{{\rm obs},i} - \mu_{\rm th}(z_i)\right)^2}{\sigma^{2}_{\mu_{\rm obs}, i}}$.

In MLE, the chosen value of $H_{0}$ is not independent of $\kappa$. That is,
one can vary either $H_{0}$ or $\kappa$, but not both. Therefore, we adopt a definition
\begin{equation}
\delta\equiv -2.5\kappa-5\log H_0+125.2\;,
\end{equation}
where $\delta$ is the ``$H_0$-free" logarithmic luminosity and
$H_0$ is in units of km $\rm s^{-1}$ $\rm Mpc^{-1}$.
With this definition, the likelihood function becomes
\begin{equation}\label{likelihood2}
\mathcal{L} = \prod_{i}
\frac{1}{\sqrt{2\pi}\,\sigma_{\mu_{\rm obs}, i}}\;\times
\exp\left[-\,\frac{\Delta_{i}^{2}}{2\sigma^{2}_{\mu_{\rm obs}, i}}\right]\;,
\end{equation}
where $\Delta_{i}=2.5\left[\alpha \log \sigma(\hb)_{i} - \log F(\hb)_{i}\right] -\delta -5\log\left[H_0\,D_L(z_i)\right]$,
with $D_L$ the luminosity distance in Mpc. The Hubble constant $H_{0}$ cancels out in Equation~(\ref{likelihood2})
when we multiply $D_L$ by $H_0$, so the constraints on the cosmological parameters are independent of the Hubble
constant. In this paper, $\alpha$ and $\delta$ are statistical ``nuisance" parameters.

To constrain the nuisance parameters and cosmological parameters simultaneously,
we use the Markov Chain Monte Carlo (MCMC) technique. Our MCMC approach generates a chain
of sample points distributed in parameter space according to the posterior probability,
using the Metropolis-Hastings algorithm with uniform prior distributions. For each
Markov chain, we generate $10^{5}$ samples based on the likelihood function. Then we
adopt the publicly available package ``triangle.py" developed by Foreman-Mackey et al. (2013)
to plot 1-D marginalized probability distributions and 2-D contours.

\section{Optimization of the model parameters}

We use the \hii~galaxies as standardizable candles and apply the
emission-line luminosity versus ionized gas velocity dispersion ($L - \sigma$) relation
(with 156 objects) to compare the standard model with the $R_{\rm h}=ct$ Universe.
In this section, we discuss how the fits have been optimized for $\Lambda$CDM, $w$CDM
and $R_{\rm h}=ct$. The outcome for each model is more fully described and discussed in subsequent sections.

\begin{figure}
\vskip-0.1in
\centerline{\includegraphics[angle=0,scale=0.6]{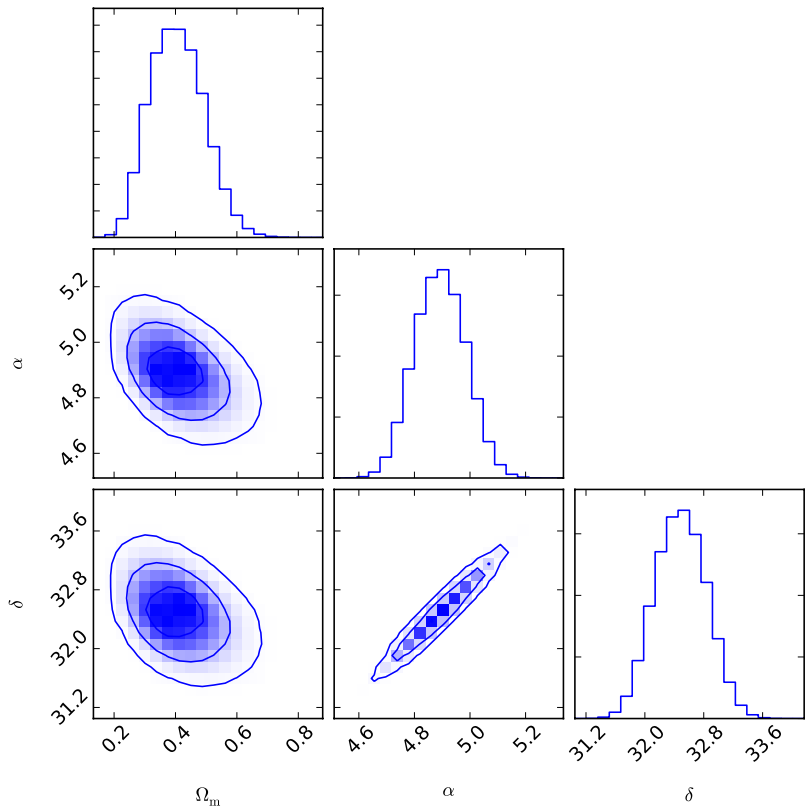}}
\caption{1-D probability distributions and 2-D regions with the $1$-$3\sigma$
contours corresponding to the parameters $\alpha$, $\delta$ and
$\Omega_{\rm m}$ in the flat $\Lambda$CDM model.}\label{LCDM}
\end{figure}

\subsection{$\Lambda$CDM}
In the most basic $\Lambda$CDM model, the dark-energy equation of state parameter,
$w_{\rm de}$, is exactly $-1$. The Hubble constant $H_0$ cancels out in Equation~(11)
when we multiply $D_L$ by $H_0$, so the essential remaining parameter in flat
$\Lambda$CDM (with $\Omega_{k}=0$) is $\Omega_{\rm m}$. The resulting constraints
on $\alpha$, $\delta$, and $\Omega_{\rm m}$ are shown in Figure~1. These contours
show that at the $1\sigma$ level, the optimized parameter values are
$\alpha=4.89^{+0.09}_{-0.09}$ $(1\sigma)$, $\delta=32.49^{+0.35}_{-0.35}$ $(1\sigma)$, and
$\Omega_{m}=0.40^{+0.09}_{-0.09}$ $(1\sigma)$. The maximum value of the joint likelihood function
for the optimized flat $\Lambda$CDM model is given by $-2\ln \mathcal{L}=563.77$,
which we shall need when comparing models using the information criteria.

\begin{figure}
\vskip-0.1in
\centerline{\includegraphics[angle=0,scale=0.5]{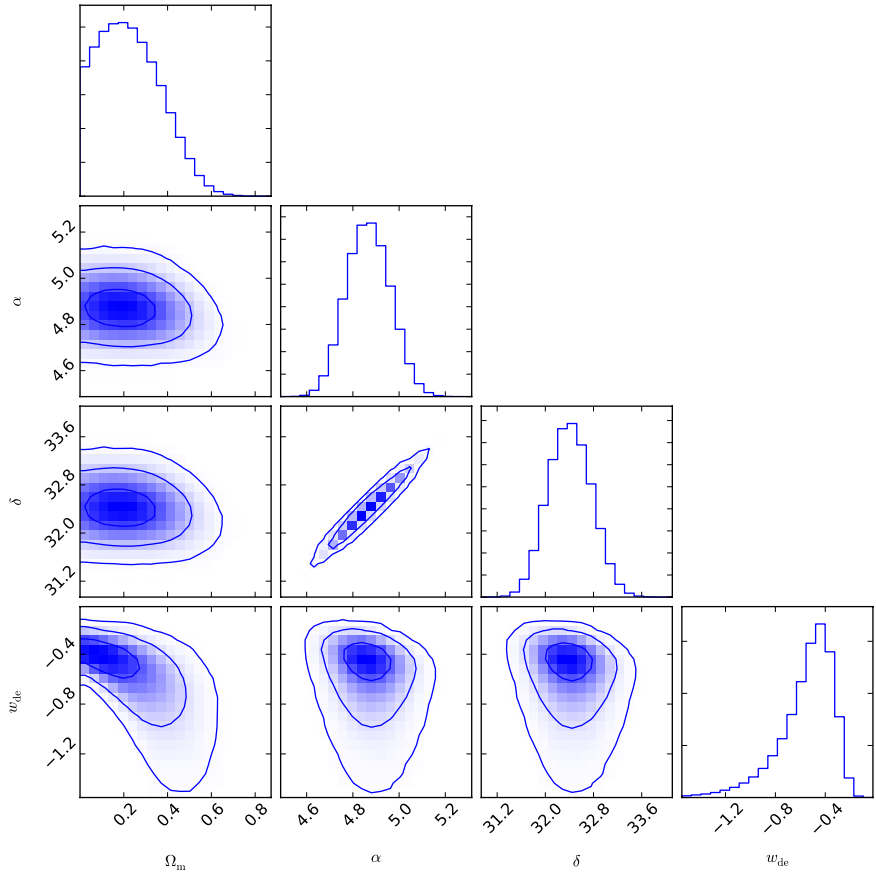}}
\caption{1-D probability distributions and 2-D regions with the $1$-$3\sigma$
contours corresponding to the parameters $\Omega_{\rm m}$, $w_{\rm de}$, $\alpha$,
and $\delta$ in the best-fit $w$CDM model.}\label{wCDM}
\end{figure}

\subsection{$w$CDM}
To allow for the greatest flexibility in this fit, we relax the assumption that
dark energy is a cosmological constant with $w_{\rm de}=-1$, and allow $w_{\rm de}$
to be free along with $\Omega_{\rm m}$. The optimized parameters
corresponding to the best-fit $w$CDM model for these 156 data are displayed in Figure~2,
which shows the 1-D probability distribution for each parameter
($\Omega_{\rm m}$, $w_{\rm de}$, $\alpha$, $\delta$), and 2-D plots of the
$1$-$3\sigma$ confidence regions for two-parameter combinations.
The best-fit values for $w$CDM are $\Omega_{\rm m}=0.22_{-0.14}^{+0.16}$ $(1\sigma)$,
$w_{\rm de}=-0.51_{-0.25}^{+0.15}$ $(1\sigma)$, $\alpha=4.87_{-0.09}^{+0.10}$ $(1\sigma)$,
and $\delta=32.40_{-0.36}^{+0.36}$ $(1\sigma)$.
The maximum value of the joint likelihood function
for the optimized $w$CDM model is given by $-2\ln \mathcal{L}=561.12$.

\begin{figure}
\vskip-0.1in
\centerline{\includegraphics[angle=0,scale=0.5]{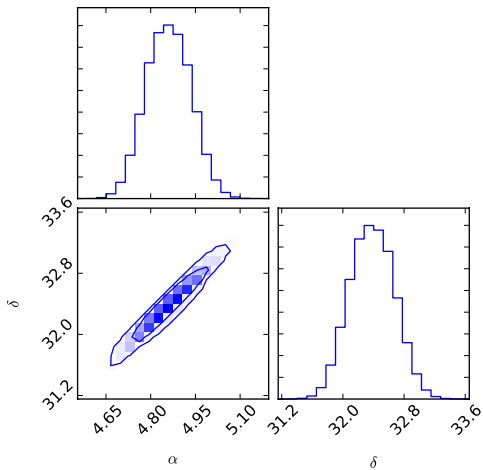}}
\caption{$1$-$3\sigma$ constraints on $\alpha$ and $\delta$ for the $R_{\rm h}=ct$ Universe.}\label{Rh}
\end{figure}

\subsection{The $R_{\rm h}=ct$ Universe}
The $R_{\rm h}=ct$ Universe has only one free parameter, $H_{0}$,
but since the Hubble constant cancels out in the product $H_{0}D_L$,
there are actually no free (model) parameters left to fit the \hii~galaxy data.
The results of fitting the $L - \sigma$ relation with this cosmology
are shown in Figure~3. We see here that the best fit corresponds to
$\alpha=4.86^{+0.08}_{-0.07}$ $(1\sigma)$ and $\delta=32.38^{+0.29}_{-0.29}$ $(1\sigma)$.
The maximum value of the joint likelihood function for the optimized
$R_{\rm h}=ct$ fit corresponds to $-2\ln \mathcal{L}=559.98$.

\section{The \hii~Galaxy Hubble Diagram}
To facilitate a direct comparison between $\Lambda$CDM and $R_{\rm h}=ct$,
we show in Figure~4 the Hubble diagrams for the combined 25 high-$z$ \hii~galaxies
and the 131 local sample (107 \hii~galaxies and 24 Giant Extragalactic \hii\ Regions).
In this figure, the observed distance moduli of 156 objects are plotted
as solid points, together with the best-fit theoretical curves (from left to right)
for the optimized flat $\Lambda$CDM model (with $\Omega_{\rm m}=0.40$,
$\alpha=4.89$, and $\delta=32.49$) and for the $R_{\rm h}=ct$ Universe
(with $\alpha=4.86$ and $\delta=32.38$). For completeness, the lower panels in Figure~4 also show
the Hubble diagram residuals relative to the best-fit cosmological models.

\begin{figure*}
\includegraphics[angle=0,scale=0.5]{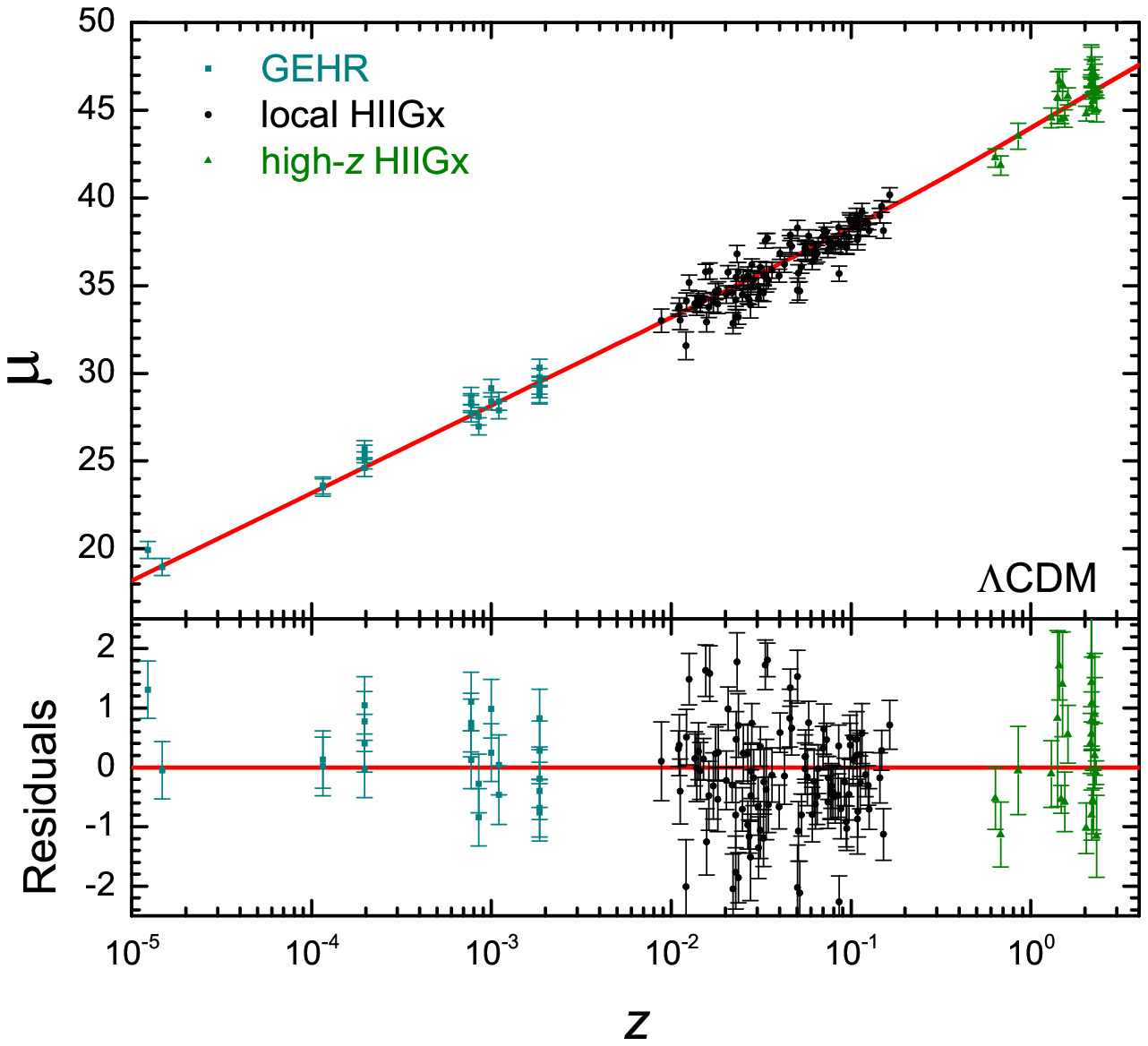} 
\includegraphics[angle=0,scale=0.5]{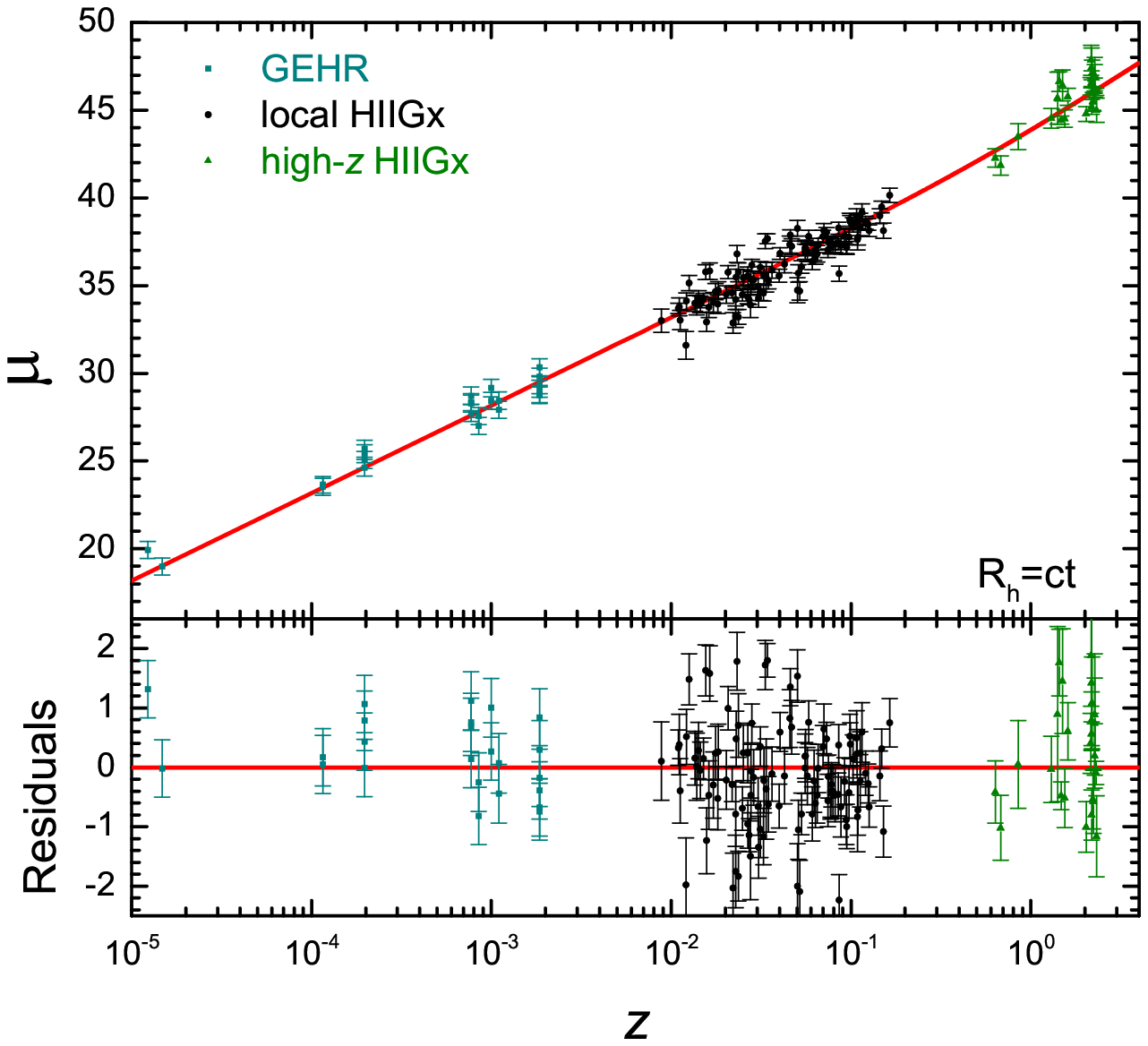}
\vskip-0.8in
    \caption{Left: Hubble diagram and Hubble diagram residuals for the 156 combined sources (including 25 high-$z$ \hii~galaxies,
    107 local \hii~galaxies, and 24 Giant Extragalactic \hii\ Regions) optimized for the flat $\Lambda$CDM model.
    Right: Same as the left panel, but now for the $R_{\rm  h}=ct$ Universe.}\label{HD}
\end{figure*}

An inspection of the Hubble diagrams in Figures~4 reveals that
both the optimized $\Lambda$CDM model and the $R_{\rm  h}=ct$ Universe
fit their respective data sets very well.
However, because these models formulate their observables (such as the
luminosity distances in Equations~\ref{DL_LCDM} and~\ref{DL_Rh}) differently, and because they
do not have the same number of free parameters, a comparison of the likelihoods
for either being closer to the `true' model must be based on model selection tools.

Several model selection tools commonly used to differentiate between cosmological models (see,
e.g., Melia \& Maier 2013, and references cited therein) include the Akaike Information
Criterion, ${\rm AIC}=-2\ln \mathcal{L}+2n$, where $n$ is the number of free parameters
(Akaike 1973; Liddle 2007; see also Burnham \& Anderson 2002, 2004),
the Kullback Information Criterion, ${\rm KIC}=-2\ln \mathcal{L}+3n$ (Cavanaugh
2004), and the Bayes Information Criterion,
${\rm BIC}=-2\ln \mathcal{L}+(\ln N)n$, where $N$ is the number of data points
(Schwarz 1978). In the case of AIC, with ${\rm AIC}_\alpha$
characterizing model $\mathcal{M}_\alpha$,
the unnormalized confidence that this model is true is the Akaike
weight $\exp(-{\rm AIC}_\alpha/2)$. Model $\mathcal{M}_\alpha$ has likelihood
\begin{equation}
P(\mathcal{M}_\alpha)= \frac{\exp(-{\rm AIC}_\alpha/2)}
{\exp(-{\rm AIC}_1/2)+\exp(-{\rm AIC}_2/2)}
\end{equation}
of being the correct choice in this one-on-one comparison. Thus, the difference
$\Delta \rm AIC \equiv {\rm AIC}_2\nobreak-{\rm AIC}_1$ determines the extent to which $\mathcal{M}_1$
is favored over~$\mathcal{M}_2$. For Kullback
and Bayes, the likelihoods are defined analogously.
In using the model selection tools,
the outcome $\Delta\equiv$ AIC$_1-$ AIC$_2$ (and analogously for KIC and BIC)
is judged `positive' in the range $\Delta=2-6$, `strong' for $\Delta=6-10$,
and `very strong' for $\Delta>10$.

With the optimized fits of the $L - \sigma$ relation (using 156 objects),
the magnitude of the difference $\Delta \rm AIC={\rm AIC}_2\nobreak-{\rm AIC}_1=5.79$, indicates that
$R_{\rm h}=ct$ (i.e. $\mathcal{M}_1$) is to be preferred over the flat $\Lambda$CDM model
(i.e. $\mathcal{M}_2$). According to Equation~(12), the likelihood of $R_{\rm h}=ct$ being
the correct choice is $P(\mathcal{M}_1)\approx 94.8\%$. For the flat $\Lambda$CDM model,
the corresponding value is $P(\mathcal{M}_2)\approx 5.2\%$. With the alternatives
KIC and BIC, the magnitude of the differences $\Delta \rm KIC={\rm KIC}_2\nobreak-{\rm KIC}_1=6.79$
and $\Delta \rm BIC={\rm BIC}_2\nobreak-{\rm BIC}_1=8.84$, indicates that $R_{\rm h}=ct$
is favored over $\Lambda$CDM by a likelihood of $\approx 96.8\%-98.8\%$ versus $1.2\%-3.2\%$.

In addition, if we relax the assumption that dark energy is a cosmological constant
with $w_{\rm de}=-1$, and allow $w_{\rm de}$ to be a free parameter along with $\Omega_{\rm m}$,
then the $w$CDM model has four free parameters (i.e., $\Omega_{\rm m}$, $w_{\rm de}$, $\alpha$,
and $\delta$), while the $R_{\rm h}=ct$ Universe has only two free parameters (i.e., $\alpha$
and $\delta$). In this case, the magnitude of the differences $\Delta \rm AIC=5.14$, $\Delta \rm KIC=7.14$,
and $\Delta \rm BIC=11.24$, indicates that $R_{\rm h}=ct$ is preferred over $w$CDM with a likelihood of
$\approx 92.9\%$ versus $7.1\%$ using AIC, $\approx 97.3\%$ versus $\approx 2.7\%$
using KIC, and $\approx 99.6\%$ versus $\approx 0.4\%$ using BIC. When the sample size
is large, as is the case here, the BIC has been shown to give more reliable results (see
Wei et al. 2015b, and references cited therein). We therefore conclude from this survey that the
current HIIGx sample is already sufficient to rule out the standard model in favor of $R_{\rm h}=ct$
at a very high confidence level ($2.5\sigma-3\sigma$).
To facilitate the comparison, we show in Table~2 the best-fit parameters for each
model, along with their $1\sigma$ uncertainties. The AIC, KIC, and BIC values are also shown in each case.

\begin{table*}
\begin{center}
{\footnotesize
\caption{Best-fitting Results in Different Cosmological Models}
\begin{tabular}{lccccccccc}
&&&&&&&& \\
\hline\hline
&&& \\
Model& $\alpha$ & $\delta$ & $\Omega_{\rm m}$ & $\Omega_{\rm de}$ & $w_{\rm de}$ & $-2\ln \mathcal{L}$ &\quad AIC&\quad KIC&\quad BIC \\
&&&&&&&& \\
\hline
&&&&&&&& \\
$R_{\rm h}=ct$            & $4.86^{+0.08}_{-0.07}$ & $32.38^{+0.29}_{-0.29}$ & --  & -- & -- & 559.98 & 563.98 & 565.98 & 570.08 \\
&&&&&&&& \\
$\Lambda$CDM              & $4.89^{+0.09}_{-0.09}$ & $32.49^{+0.35}_{-0.35}$ & $0.40^{+0.09}_{-0.09}$ & $1.0-\Omega_{\rm m}$ & $-1$(fixed) & 563.77 & 569.77 & 572.77 & 578.92 \\
&&&&&&&& \\
$w$CDM                    & $4.87^{+0.10}_{-0.09}$ & $32.40^{+0.36}_{-0.36}$ & $0.22^{+0.16}_{-0.14}$ & $1.0-\Omega_{\rm m}$ & $-0.51^{+0.15}_{-0.25}$ & 561.12 & 569.12 & 573.12 & 581.32 \\
&&&&&&&& \\
\hline\hline
\end{tabular}
}
\end{center}
\end{table*}

A possible concern with this analysis is the fact that the redshifts of 24 GEHR
were calculated by Terlevich et al. (2015) from the measured distance moduli using as prior the value
of $H_{0}$ reported by Ch{\'a}vez et al. (2012). This may induce a subtle dependency with that value
of $H_{0}$, even though our methodology tries to avoid this. To test whether our conclusions are affected
in this way, we repeat the analysis by removing the GEHR subsample. The new results are presented in Table~3.
We find that $R_{\rm h}=ct$ is preferred over the flat $\Lambda$CDM model with a likelihood of
$\approx 92.8\%$ versus $7.2\%$ using AIC, $\approx 95.5\%$ versus $\approx 4.5\%$
using KIC, and $\approx 98.2\%$ versus $\approx 1.8\%$ using BIC; and $R_{\rm h}=ct$
is preferred over $w$CDM with a likelihood of $\approx 90.4\%$ versus $9.6\%$ using AIC,
$\approx 96.2\%$ versus $\approx 3.8\%$ using KIC, and $\approx 99.4\%$ versus $\approx 0.6\%$ using BIC.
By comparing these outcomes with the results of the full sample, we conclude that even if the GEHR
subsample is removed, one still gets a very similar outcome to the analysis using the full source catalog.

\begin{table*}
\begin{center}
{\footnotesize
\caption{Best-fitting Results in Different Cosmological Models when the subsample of 24 GEHR is removed}
\begin{tabular}{lccccccccc}
&&&&&&&& \\
\hline\hline
&&& \\
Model& $\alpha$ & $\delta$ & $\Omega_{\rm m}$ & $\Omega_{\rm de}$ & $w_{\rm de}$ & $-2\ln \mathcal{L}$ &\quad AIC&\quad KIC&\quad BIC \\
&&&&&&&& \\
\hline
&&&&&&&& \\
$R_{\rm h}=ct$            & $4.60^{+0.10}_{-0.10}$ & $31.31^{+0.41}_{-0.41}$ & --  & -- & -- & 501.44 & 505.44 & 507.44 & 511.21 \\
&&&&&&&& \\
$\Lambda$CDM              & $4.56^{+0.13}_{-0.13}$ & $31.16^{+0.50}_{-0.50}$ & $0.50^{+0.11}_{-0.10}$ & $1.0-\Omega_{\rm m}$ & $-1$(fixed) & 504.56 & 510.56 & 513.56 & 519.21 \\
&&&&&&&& \\
$w$CDM                    & $4.54^{+0.13}_{-0.13}$ & $31.09^{+0.52}_{-0.51}$ & $0.30^{+0.19}_{-0.18}$ & $1.0-\Omega_{\rm m}$ & $-0.45^{+0.16}_{-0.27}$ & 501.92 & 509.92 & 513.92 & 521.45 \\
&&&&&&&& \\
\hline\hline
\end{tabular}
}
\end{center}
\end{table*}

\section{Discussion and Conclusions}
HIIGx and GEHR have been proposed as useful standard candles due to the correlation between
their velocity dispersion and the luminosity of their $\hb$ emission line.
Given that HIIGx can be observed to $z\sim3$, they constitute a promising
new cosmic tracer which may allow us to obtain better constraints on cosmological parameters
than those currently available using tracers at lower redshifts.
In this paper, we have added some support to the argument that HIIGx and GEHR can eventually be
used to carry out stringent tests on various cosmological models.

We have confirmed the notion advanced previously that examining the
correlation between their velocity dispersion and the $\hb$-line luminosity can indeed produce
a luminosity indicator with sufficient reliability to study the expansion of the Universe.
We have used the sample of 25 high-redshift HIIGx, 107 local HIIGx, and 24 GEHR to
test and compare the standard model $\Lambda$CDM and the $R_{\rm h}=ct$ Universe.
We have individually optimized the parameters in each model by maximizing the likelihood
function. In this regard, we emphasize that one should always optimize parameters by carrying out a
maximum likelihood estimation in any situation where the error in the observed distance modulus
$\sigma_{\mu_{\rm obs}}$ is dependent on the value of one or more free parameters. It is
not correct in such circumstances to simply rely on a $\chi^2$ minimization.
In the flat $\Lambda$CDM model, the resulting Hubble diagram leads to a best-fit value
$\Omega_{\rm m}=0.40_{-0.09}^{+0.09}$. A statistically acceptable fit
to the Hubble diagram with $w$CDM (the version of $\Lambda$CDM with a dark-energy equation of
state $w_{\rm de}\equiv p_{\rm de}/\rho_{\rm de}$ rather than $w_{\rm de}=w_{\Lambda}=-1$)
is possible only with $\Omega_{\rm m}=0.22_{-0.14}^{+0.16}$ and $w_{\rm de}=-0.51_{-0.25}^{+0.15}$.
These values, however, are not fully consistent with the concordance model.

More importantly, we have found that, when the parameter optimization is handled via
maximum likelihood optimization, the Akaike, Kullback and Bayes Information
Criteria tend to favor the $R_{\rm h}=ct$ Universe. Since the flat $\Lambda$CDM model
(with $\Omega_{\rm m}$, $\alpha$, and $\delta$) has one more free parameter than
$R_{\rm h}=ct$ (i.e., $\alpha$ and $\delta$), the latter is preferred over the former
with a likelihood of $\approx 94.8\%$ versus $\approx 5.2\%$ using AIC, $\approx 96.8\%$
versus $\approx 3.2\%$ using KIC, and $\approx 98.8\%$ versus $\approx 1.2\%$
using BIC. If we relax the assumption that dark energy is a cosmological constant
with $w_{\rm de}=-1$, and allow $w_{\rm de}$ to be a free parameter along with $\Omega_{\rm m}$,
then the $w$CDM model has four free parameters (i.e., $\Omega_{\rm m}$, $w_{\rm de}$, $\alpha$,
and $\delta$). We find that $R_{\rm h}=ct$ is preferred over $w$CDM with a likelihood of
$\approx 92.9\%$ versus $7.1\%$ using AIC, $\approx 97.3\%$ versus $\approx 2.7\%$
using KIC, and $\approx 99.6\%$ versus $\approx 0.4\%$ using BIC.

In other words, the current HIIGx sample is sufficient to rule out the standard model in favor of $R_{\rm h}=ct$
at a very high confidence level. The consequences of this important result are being explored
elsewhere, including the growing possibility that inflation may have been unnecessary
to resolve any perceived `horizon problem' and therefore may have simply never
happened (Melia 2013a).

We close this discussion by pointing out three important caveats to our conclusions.
First, since the cosmological parameters are more sensitive to the high-$z$ observational
data than the low-$z$ ones, most of the weight of the constraints obtained in this work
(e.g., for the parameter of the equation of state of dark energy) is from the high-$z$
sample of only 25 HIIGx. Secondly, the systematic uncertainties of
the $L(\hb)-\sigma$ correlation need to be better understood, which may affect HII galaxies
as cosmological probes. The associated systematic uncertainties include the size of the burst,
the age of the burst, the oxygen abundance of HII galaxies, and the internal extinction correction
(Ch{\'a}vez et al. 2016).

Some progress has already been made with attempts at mitigating these
uncertainties (Melnick et al. 1988; Ch\'avez et al. 2014), though efforts such as
these also highlight the need to probe all possible sources of systematic errors
more deeply. For example, an important consideration is the exclusion of rotationally
supported systems, which clearly would skew the $L(H\beta)-\sigma$ relation.
Melnick et al. (1988) and Ch\'avez et al. (2014, 2016) have proposed using an
upper limit to the velocity dispersion of $\log\sigma(H\beta)\sim 1.8$ km s$^{-1}$
to minimize this possibility, though at a significant cost---of greatly reducing the
catalog of suitable sources. Nonetheless, even with this limit, there is no
guarantee that such a systematic effect is completely removed. As a second
example, since the $L(H\beta)-\sigma$ relation is essentially a correlation
between the ionizing flux produced by the massive stars, and the velocity
field in the potential well due to stars and gas, any systematic variation of
the initial mass function will affect the mass-luminosity ratio and therefore the
slope and/or zero point of the relation (Ch\'avez et al. 2014).

A third important caveat is that our constraints are somewhat weaker than the results of
some other cosmological probes, and they have uncertainties because of the small HIIGx
sample effect. To increase the significance of the constraints, one needs a larger sample.
Fortunately, with the help of current facilities, such as the K-band Multi Object Spectrograph
at the Very Large Telescope, a larger sample of high-$z$ HIIGx with high quality data
will be observed in the near future, which will provide much better and competitive
constraints on the cosmological parameters (Terlevich et al. 2015).

\section*{Acknowledgments}
We thank the anonymous referee for insightful comments that have helped us improve
the presentation of the paper.
This work is partially supported by the National Basic Research Program (``973" Program)
of China (Grants 2014CB845800 and 2013CB834900), the National Natural Science Foundation
of China (grants Nos. 11322328 and 11373068), the Youth Innovation Promotion Association
(2011231), the Strategic Priority Research Program ``The Emergence of Cosmological
Structures" (Grant No. XDB09000000) of the Chinese Academy of Sciences,
the Natural Science Foundation of Jiangsu Province (Grant No. BK20161096),
and the Guangxi Key Laboratory for Relativistic Astrophysics.
F.M. is grateful to Amherst College for its support through a John Woodruff Simpson
Lectureship, and to Purple Mountain Observatory in Nanjing, China, for its hospitality
while part of this work was being carried out. This work was partially supported by grant
2012T1J0011 from The Chinese Academy of Sciences Visiting Professorships for Senior
International Scientists, and grant GDJ20120491013 from the Chinese State Administration
of Foreign Experts Affairs.

\end{document}